\documentclass[12pt]{article}
\pdfoutput=1
\usepackage{amsmath}
\usepackage{amsfonts}
\usepackage{setspace}
\usepackage{subcaption}
\usepackage{enumerate}
\usepackage{graphicx}
\usepackage[hidelinks]{hyperref} 
\numberwithin{equation}{section}
\usepackage[utf8]{inputenc}
\newcommand{\gsim}{\lower.7ex\hbox{$\;\stackrel{\textstyle>}{\sim}\;$}}
\newcommand{\lsim}{\lower.7ex\hbox{$\;\stackrel{\textstyle<}{\sim}\;$}}
\def\O{{\mathcal O}}

\newcommand{\be}{\begin{equation}}
\newcommand{\ee}{\end{equation}}
\newcommand{\bea}{\begin{eqnarray}}
\newcommand{\eea}{\end{eqnarray}}

\newcommand{\comment}[1]{}
\newcommand{\expect}[1]{\left\langle #1 \right\rangle}

\newcommand{\bsb}{\boldsymbol}
\newcommand{\Cint}{C\kern-1em\int}

\def\ep{\epsilon}

\def\d{\partial}
\def\vphi{\varphi}
\def\x{{\vec x}}

\def\O{\mathcal{O}}

\def\r{{\bsb r}}

\def\tr{{\rm Tr}}

\def\arctanh{{\rm arctanh}}
\def\bphi{{\bsb \vphi}}
\def\Im{\rm Im ~}

\begin{document}
\vspace*{-1. cm}
\begin{center}
{\bf \Large Infrared dynamics of a light scalar field in de Sitter}
\vskip 1cm
{{\bf Mehrdad Mirbabayi} }
\vskip 0.5cm
{\normalsize {\em International Centre for Theoretical Physics, Trieste, Italy}}
\vskip 0.2cm
{\normalsize {\em Stanford Institute for Theoretical Physics, Stanford University,\\ Stanford, CA 94305, USA}}
\end{center}
\vspace{.8cm}
{\noindent \textbf{Abstract:}  
Inertial observers in de Sitter are surrounded by a horizon and see thermal fluctuations. To them, a massless scalar field appears to follow a random motion but any attractive potential, no matter how weak, will eventually stabilize the field. We study this {\em thermalization} process in the static patch (the spacetime region accessible to an individual observer) via a truncation to the low frequency spectrum. We focus on the distribution of the field averaged over a subhorizon region. At timescales much longer than the inverse temperature and to leading order in the coupling, we find the evolution to be Markovian, governed by the same Fokker-Planck equation that arises when the theory is studied in the inflationary setup.

\vspace{0.3cm}
\vspace{-1cm}
\vskip 1cm
\section{Introduction}
To an inertial observer de Sitter spacetime appears to have a horizon. The accessible part of the spacetime is called the static patch and is described by the metric
\be\label{static}
ds^2=-(1- r^2)dt^2+\frac{dr^2}{1- r^2}+r^2 d\Omega^2,
\ee
where we have set the radius of curvature to $1$ and $d\Omega^2$ is the metric of a unit 2-sphere. Suppose there is a light scalar field $\phi$ in this spacetime. A natural question the observer can ask is how an initial perturbation in $\phi$ relaxes. For a free massive field, the late time behavior is controlled by the ``quasinormal modes'' of the Klein-Gordon equation on metric \eqref{static}:
\be
(\Box_{\rm dS} -m^2)\phi=0.
\ee
The quasinormal modes are defined by making use of the time-independence of the metric and going to the frequency space. Then, one looks for the eigenmodes of the resulting Schr\"odinger-like equation that are regular at $r=0$ and outgoing at the horizon, as appropriate for an initial value problem. They have complex frequencies, corresponding to the fact that perturbations decay in time. The late-time decay is dominated by the mode with the smallest $|\Im \omega|$. It is spherically symmetric ($l=0$), and has a purely imaginary frequency, given in the $m\ll 1$ limit by
\be\label{omega0}
\omega_0 \simeq -i\frac{m^2}{3 }.
\ee
Our goal here is to ask the same question when the mass term is replaced by a general attractive potential $V(\phi)$. This is not a problem of finding small corrections to \eqref{omega0}. Due to the thermal fluctuations, a light field typically has large excursions $\phi\gg 1$. So it is meaningful only to ask about the relaxation of perturbations $\delta\phi\gg 1$, for which the nonlinearities can be essential even at weak coupling. 

The state describing observables in the static patch is a mixed state, which can be thought of as arising from tracing over the degrees of freedom behind the horizon. At equilibrium this is a thermal state (also known as the Hartle-Hawking state). The corresponding density matrix is given by a path integral over a cut 4-sphere, which is the Euclidean continuation of dS$_4$. This is reviewed in section \ref{sec:HH}. However the problem we formulated above requires perturbing this thermal state. Simple observables, such as the field smeared over a finite region as in figure \ref{fig:slice}-Left, gradually lose the memory of the perturbation and relax to their thermal distribution. 

In section \ref{sec:simple} we will discuss these simple observables and their reduced density matrix. We will see that if the size of the region over which the field is smeared is $r_0\sim 1$, the off-diagonal elements of this density matrix are small. We can then focus on the diagonal, which is regarded as an approximately classical probability distribution $p(t,\vphi)$. At equilibrium, a saddle point approximation to the above-mentioned path integral over a 4-sphere gives a very simple answer:
\be\label{eq}
p_{\rm eq}(\vphi) \sim e^{-S_4 V(\vphi)},
\ee
where $S_4= 8\pi^2/3$ is the 4-volume of a unit 4-sphere. 

In section \ref{sec:FP} we study deviation from this thermal equilibrium. One can perturb the state by inserting operators in the path integral over the 4-sphere. A Lorentzian analog would be to ask how the density matrix evolves after performing a simple measurement and registering $\vphi_B$. The diagonal part of the reduced density matrix now defines a conditional probability $p(t,\vphi|\vphi_B)$. Its evolution, which is generated by a time-independent Hamiltonian, simplifies in the long time limit when one can set up a systematic low frequency expansion. At leading order we find
\be\label{FP}
\d_t p(t,\vphi|\vphi_B) = \frac{1}{8\pi^2}\d_\vphi^2 p(t,\vphi|\vphi_B) +\frac{1}{3}\d_\vphi(V'(\vphi) p(t,\vphi|\vphi_B)).
\ee
This is a Markovian evolution, i.e. an evolution that depends only on the current state and not the history. The eigenmodes of this ``Fokker-Planck'' equation control the relaxation of the system to the equilibrium state \eqref{eq}, and their (pure imaginary) frequencies will give the relaxation time we are interested in. We will conclude in section \ref{sec:con} with a discussion of higher order corrections in the low energy expansion.
\begin{figure}[t]
\centering
\includegraphics[scale =1.]{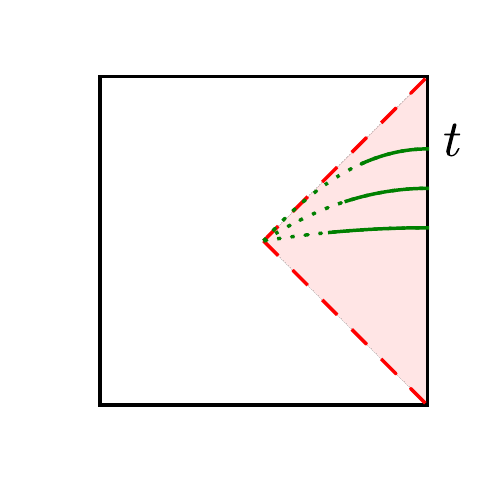} 
~~~~~~~~~~~~~~\includegraphics[scale =1.]{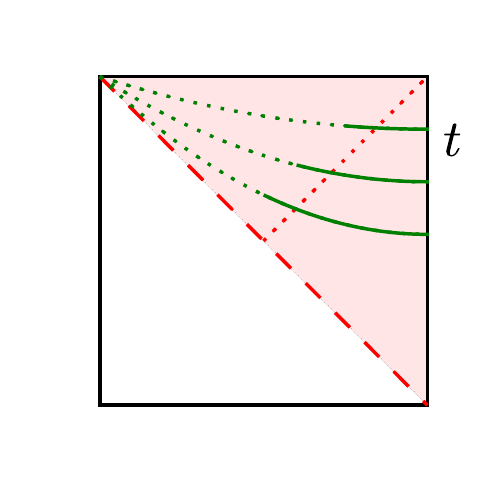} 
\caption{\small{Left: We study the evolution of $\phi$ smeared over a subhorizon region in the static patch (shaded). Right: In the stochastic approach of Starobinsky and Yokoyama the field is smeared over a superhorizon region of fixed physical size.}}
\label{fig:slice}
\end{figure}

Before embarking on the static patch analysis, it is worth pointing out another natural question in this field theory. This time it is asked by a dS meta-observer who has access to the cosmological data. This observer describes the spacetime by the metric
\be\label{flat}
ds^2= -dt^2 + e^{2t} d\x^2,
\ee
and the question is how fluctuations of $\phi$ correlate at superhorizon separation (i.e. when $e^t |\Delta \x|\gg 1$). This is the type of question commonly studied in the context of inflation, for instance to infer predictions for the CMB correlation functions. This second question has been studied extensively starting with the pioneering works by Starobinsky \cite{Starobinsky}, and Starobinsky and Yokoyama \cite{SY}. More recent works include \cite{Tsamis,Finelli,Finelli2,Burgess,Vennin,Kitamoto,Markkanen,Lopez,Gorbenko}. The approach of \cite{Gorbenko} has been particularly inspiring to the present work. 

Nonlinearities are essential also in this formulation of the problem. Perturbative calculation of correlation functions would fail here since the interactions do not shut off when the modes exit the horizon. For instance, in $\lambda \phi^4$ theory the loop expansion of the momentum space correlators with characteristic momentum $k$ becomes an expansion in powers of $\lambda (t-t_k)^2$ and breaks down when $t$ is sufficiently larger than the horizon crossing time $t_k=\log k$.\footnote{As emphasized in \cite{Marolf} the power-law growth in $t$ will saturate if there is a nonzero mass. However as follows from \eqref{omega0} the expansion parameter is $\lambda/m^4$ in this case. Therefore the perturbative expansion still breaks down for small enough mass.} The stochastic method of \cite{Starobinsky,SY} can be thought of as a clever way of resumming the leading diagrams. 

The way this resummation is achieved in \cite{Starobinsky,SY} is through the study of the 1-patch distribution function $\rho_1(t,\vphi)$ defined as the distribution function of $\phi_\ep(t)$, the average field over a superhorizon region of fixed physical size $\sim 1/\ep$ (see figure \ref{fig:slice}-Right). This is the diagonal element of the reduced density matrix for the single observable $\phi_\ep$, and it satisfies the same Fokker-Planck equation \eqref{FP}.
Cosmological observables are calculated in \cite{SY} based on this machinery.

Curiously, what we called $p(t,\vphi|\vphi_B)$ in the first formulation of the question closely resembles $\rho_1$ of \cite{Starobinsky,SY}. The difference is that in the former the field has to be averaged over a subhorizon region which is accessible to the dS observer. The fact that $p$ and $\rho_1$ satisfy the same Fokker-Planck equation is not a coincident. The eigenvalues of this equation $\omega_i$ determine the long time (or long distance) behavior of the correlators. This should not depend on the details of the experimental setup. Hence we are merely proposing a new perspective into the same problem, besides setting up a perturbation theory that, at least in principle, allows a systematic inclusion of subleading corrections.
\section{The state in equilibrium}\label{sec:HH}
The vacuum wavefunction in global de Sitter spacetime can be defined using the Hartle-Hawking prescription \cite{HH}. We will review the construction in this section (a similar discussion can be found in \cite{Higuchi}). First we cut the global manifold, described by the metric
\be\label{closed}
ds^2 = -d\tau^2 +  \cosh^2(\tau) (d\chi^2 + \sin^2\chi d\Omega^2),
\ee
along a minimal 3-sphere $S^3$, say at $\tau=0$. This $S^3$ slice bounds a Euclidean 4-hemisphere $S^4_-$ of radius $1$ (see figure \ref{fig:penrose}).
\begin{figure}[t]
\centering
\includegraphics[scale =1.3]{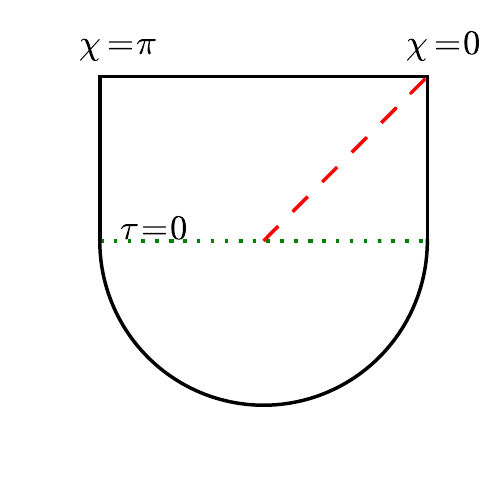} 
\caption{\small{The Penrose diagram for de Sitter spacetime connected to a Euclidean hemisphere along the $\tau=0$ slice (the dotted line). A 2-sphere is suppressed. Its radius, given by $\sin\chi$, expands and then shrinks from the right (north pole) to the left (south pole). The future horizon of an inertial observer at the north pole is shown with a red dashed line.}}
\label{fig:penrose}
\end{figure}}
The wavefuction for any configuration $\bphi$, where $\bphi:S^3\to \mathbb{R}$ is a specified function on this slice, is given by the Euclidean path integral over the hemisphere with the boundary condition $\bphi$:
\be
\Psi_{HH}[\bphi]=\int^{\bphi} D\phi e^{-S_E[\phi]},\qquad \text{over}\quad S^4_-
\ee
where an overall normalization constant has been absorbed in the definition of the path integral measure. The state of the universe at later times is determined by the unitary evolution corresponding to the global time Hamiltonian, i.e. the generator of $\tau$ translations. When restricted to large spherical harmonics on $S^3$, the closed topology of the spatial slices becomes irrelevant and this prescription coincides with the adiabatic (or Bunch-Davies) vacuum choice made in the inflationary calculations on metric \eqref{flat}. 

\begin{figure}[t]
\centering
\includegraphics[scale =1.3]{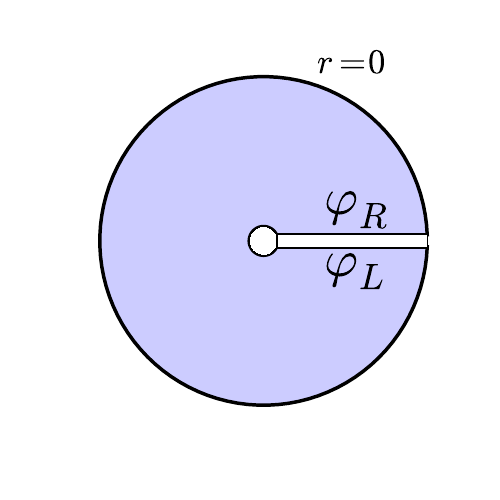} 
\caption{\small{The Hartle-Hawking state for the static patch is given by the path integral over a cut 4-sphere (the shaded region). A 2-sphere is suppressed. Its radius is $1$ at the center, but it shrinks to $0$ and caps off the geometry at the outer edge. The boundary conditions imposed on the two sides of the cut determine the arguments of the density matrix $\rho_{HH}[\bphi_L,\bphi_R]$.}}
\label{fig:rho}
\end{figure}

The observers in the static patch \eqref{static} have access to half of the initial $S^3$ slice, say $0\leq\chi<\frac{\pi}{2}$. Their state can be obtained by tracing over the other half. The result is a mixed state, described by a density matrix 
\be
{\rho}_{HH} =\tr_{\frac{\pi}{2}<\chi\leq \pi} {\Psi}_{HH}
{\Psi}_{HH}^\dagger.
\ee
More explicitly, ${\rho}$ is a functional of two field configurations $\bphi_L$ and $\bphi_R$, where now $\bphi_{L/R}:S^3_-\to \mathbb{R}$. The density matrix is given by a path integral over a cut $S^4$: one 4-hemisphere for $\Psi$ and another for $\Psi^\dagger$, glued together along $\frac{\pi}{2}<\chi\leq \pi$. There is a cut along $0\leq \chi <\frac{\pi}{2}$ with boundary conditions $\bphi_L,\bphi_R$ imposed on the two sides (see figure \ref{fig:rho})
\be\label{rho}
\rho_{HH}[\bphi_L,\bphi_R]=\int_{\bphi_L}^{\bphi_R} D\phi e^{-S_E[\phi]},\qquad 
\text{over cut $S^4$}.
\ee
This density matrix describes a thermal state with temperature $T=\frac{1}{2\pi}$ \cite{GH}:
\be\label{th}
{\rho}_{HH} = \frac{1}{Z} \sum_n e^{-2\pi E_n} |E_n\rangle\langle E_n|.
\ee
This can be seen as follows. When analytically continued to Euclidean time $t\to i t_E$, the metric \eqref{static} becomes the metric of a 4-sphere. $t$ translation which is an isometry of dS becomes a rotation on the 4-sphere. $t_E$ is the corresponding angular variable with the fixed point at $r=1$ (i.e. the horizon in Lorentzian signature). The Hartle-Hawking prescription fixes the periodicity $t_E\sim t_E+2\pi$, thereby ensuring that there is no conical singularity at $r=1$. The static-patch Hamiltonian $H$ is the generator of $t$ translations. Hence the Euclidean path integral \eqref{rho} can be sliced along constant $t_E$ hypersurfaces and interpreted as the matrix elements of the evolution operator (up to a normalization constant)
\be
\rho_{HH}\propto U(-2\pi i)=e^{-2\pi H}.
\ee
Using a complete set of energy eigenstates one arrives at \eqref{th}.

Once the state at $t=0$ is given, the state of the system for all $t$ can be determined by evolving with $H$:
\be\label{evolve}
{\rho}(t) = e^{-iHt} {\rho}(0) e^{iHt}.
\ee
In other words, the domain of dependence of the region $0\leq \chi < \frac{\pi}{2}$ on the initial $S^3$ hypersurface is the entire static patch. The thermal state we found above does not change under \eqref{evolve}. That is to say, it is the equilibrium state.
\section{The state of simple observables}\label{sec:simple}
The question we formulated in the introduction could in principle be answered in this setup by calculating the Euclidean correlators as a function of $t_E$ and analytically continuing them to large real time separation. However, this requires exponential accuracy in the Euclidean calculation, and hence it is not very practical. But see \cite{Marolf,Lopez} for related works. 

To study the approach to equilibrium, we instead study the real time evolution of a perturbed state. Clearly no such state ${\rho}(t=0)\neq {\rho}_{HH}$ can relax to the thermal Hartle-Hawking state under the evolution \eqref{evolve}. Thermalization is an approximate notion that emerges when we reduce the density matrix to a subset of local observables. Intuitively it happens because perturbations move toward the horizon and eventually leave any region of interest in the interior.

We consider as our simple observable the spatial average of $\phi$ over a constant-$t$ slice, weighted by a spherically symmetric function $w:S^3_-\to \mathbb{R}$ with characteristic size $r_0$. We denote this by $\bar \phi$:
\be\label{phibar}
\bar \phi(t) \equiv\int d^3\r\ w(r)\phi(t,\r)\equiv \int_0^{1} \frac{dr r^2}{\sqrt{1-r^2}}w(r)\int d\hat r\ \phi(t,\r).
\ee
where $d^3\r$ is the measure on the 3-hemisphere $S^3_-$ and $w$ is normalized
\be
\int d^3\r w(r) = 1.
\ee
The reduced density matrix is now defined for a single degree of freedom: $\rho(t,\vphi_L,\vphi_R)$. It is obtained from $\rho[t,\bphi_L,\bphi_R]$ by tracing over all field configurations with fixed $\bar\bphi_L = \vphi_L$ and $\bar\bphi_R =\vphi_R$. This can be implemented by introducing sources:
\be\label{reduce}
\rho(t,\vphi_L,\vphi_R)=\int \frac{dj_L dj_R}{(2\pi)^2}e^{i(j_L\vphi_L+j_R\vphi_R)}\ Z(t,j_L,j_R),
\ee
where the reduced partition function is related to the partition functional $Z[t,J_L,J_R]$ (with  $J_{L/R}:S^3_-\to \mathbb{R}$ source configurations on the spatial slice) via a restriction of the sources:
\be
Z(t,j_L,j_R)\equiv \int D J_L D J_R\ Z[t,J_L,J_R]\delta(J_L-j_L w)
\delta(J_R-j_R w),
\ee
and the partition functional is given by
\be
Z[t,J_L,J_R]\equiv \int D\bphi_L D\bphi_R
 e^{-i\int {d^3\r}[J_L(\r)\bphi_L(\r)+J_R(\r)\bphi_R(\r)]}
\rho[t,\bphi_L,\bphi_R].
\ee

\begin{figure}[t]
\centering
\includegraphics[scale =1.]{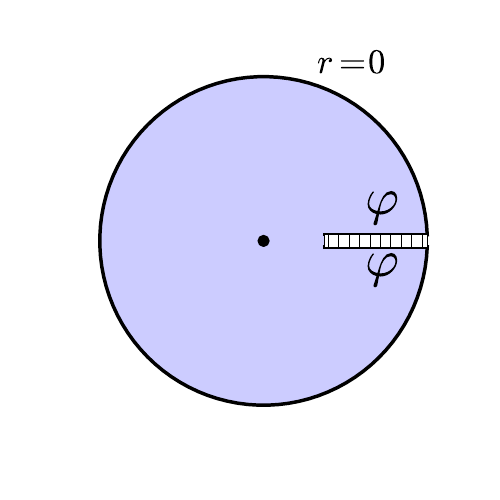} 
~~~~~~~~~~~~\includegraphics[scale =1.]{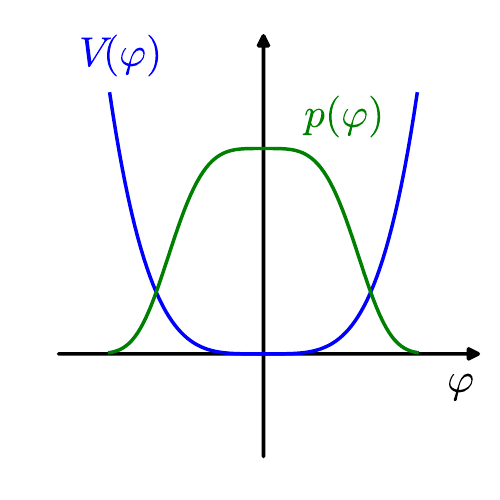} 
\caption{\small{Left: $p_{HH}(\vphi)$ is obtained by gluing back all degrees of freedom on the two sides of the cut except for the average field $\bar\phi$ which is set to $\vphi$ on both sides. Right: The equilibrium distribution of $\bar\phi$. }}
\label{fig:red}
\end{figure}

When applied to the Hartle-Hawking state \eqref{rho}, this tracing has the geometric meaning of identifying everything except $\bar\phi$ on the two sides of the cut. Let us consider the diagonal element 
\be
p_{HH}(\vphi)\equiv\rho_{HH}(\vphi,\vphi).
\ee
It is given by the path integral in figure \ref{fig:red}-Left. If $V(\phi)=0$, a saddle point of the path integral with this boundary condition would be the uniform solution $\phi=\vphi$ over the entire $S^4$. On this solution, the Euclidean action $S_E=0$ which is the absolute minimum. So this is the leading saddle. The vanishing of the action is the consequence of the fact that there is no preferred value for a shift-symmetric field.

If the potential $V(\phi)$ is nontrivial, then a constant $\phi$ will not be a solution over the 4-sphere. However, the gradient of the true solution would be small if the potential is not too steep. Hence the classical solution would be approximately uniform $\phi \simeq \vphi$ and we get 
\be\label{eq2}
p_{HH}(\vphi) = A e^{-\frac{8\pi^2}{3} V(\vphi)(1 + \O(\ep,\eta))},
\ee
where $A$ is the normalization factor and the ``slow-roll'' parameters
\be\label{sr}
\ep \equiv \frac{{V'}^2}{V}\ll 1,\qquad \eta \equiv \frac{V''}{V},
\ee
control, respectively, the gradient corrections and the one-loop determinant. $p_{HH}$ is the well-known equilibrium distribution that was originally obtained using stochastic methods \cite{Starobinsky,SY}. The attractive potential contains the Brownian spread of the field. See figure \ref{fig:red}-Right.

Note that we did not need to specify the size $r_0$ of the region over which the field is averaged. The characteristics of the weight function $w$ start showing up at next-to-leading order in the slow-roll expansion. However, $r_0$ has the more important effect of controlling the off-diagonal elements of the reduced density matrix. One can estimate that $\rho(\vphi_L,\vphi_R)$ is appreciable over a range
\be
|\vphi_L-\vphi_R|\sim \frac{1}{r_0}.
\ee
Hence to have a notion of classical history of $\bar\phi$ with resolution $\Delta\vphi\sim 1 (\ll$ width of $p_{HH}$), we must choose $r_0\sim 1$. This is qualitatively similar to smearing operators over a timescale of order $1$. 
\section{The state out of equilibrium}\label{sec:FP}
Suppose $\bar\phi$ is measured to be $\vphi_B$ at $t=0$. This collapses the Hartle-Hawking state in the $\bar\phi$ subspace to $p(t=0,\vphi)=\delta(\vphi-\vphi_B)$. Finding the subsequent evolution of this excited state as a function of $\vphi_B$ would then allow us to calculate the correlator $\expect{\bar\phi(t)\bar\phi(0)}$. Hence, even though this is a very particular perturbed state, obtained by applying the projection operator at time $0$, it is sufficient to answer the question about relaxation. Moreover, we will see that the evolution quickly becomes Markovian in the $\bar\phi$ subspace and forgets the choice of the initial state. 

\begin{figure}[t]
\centering
\includegraphics[scale =1.3]{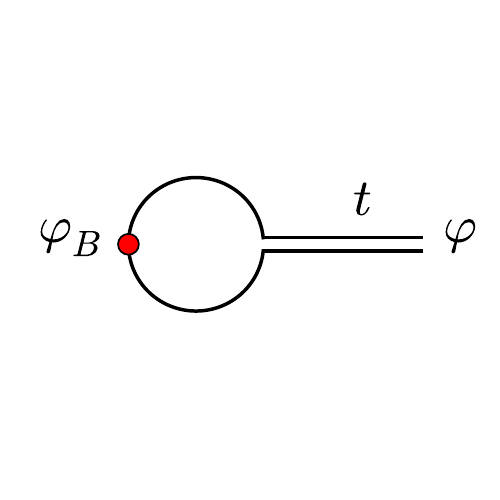} 
\caption{\small{We consider a perturbed state obtained by a projection into $\bar\phi=\vphi_B$ (the red dot) at $t=i\pi$. Hence the path integral for the diagonal element of the reduced density matrix $p(t,\vphi|\vphi_B)$ is symmetric. It consists of a sum over the product of matrix elements of $U(-i\pi+t)$ and $U^{-1}(i\pi+t)$.}}
\label{fig:p}
\end{figure}

It simplifies the algebra to symmetrize the insertions along the thermal circle, i.e. to apply the second projection after $-i\pi+t$ (see figure \ref{fig:p}).
So we are calculating
\be\label{p}
p(t,\vphi|\vphi_B)=\int \frac{dj  dk}{(2\pi)^2} e^{i(j\vphi+k\vphi_B)}Z(t,j|k)
\ee
where in terms of the weight function $w(r)$
\be\label{Z}\begin{split}
Z(t,j|k) &= \int DJ DK  Z[t,J|K] \delta(J-j w)\delta(K-kw) \\[10pt]
Z[t,J|K]&=\int D\bphi_B D\bphi
e^{-i \int d^3\r (J(\r)\bphi(\r)+ K(\r)\bphi_B(\r))}
\Psi^\dagger[t,\bphi|\bphi_B]\Psi[t,\bphi|\bphi_B],\\[10pt]
\Psi[t,\bphi|\bphi_B]& = \langle \bphi|e^{-i H (t-i\pi)}|\bphi_B\rangle,\\[10pt]
\Psi^\dagger[t,\bphi|\bphi_B]& = \langle \bphi_B|e^{i H (t+i\pi)}|\bphi\rangle.
\end{split}\ee
$\Psi$ has an expression in terms of a path integral with boundary condition $\bphi_B$ imposed on one side, evolved for $\pi$ in Euclidean time and then Wick rotated to real time and evolved for time $t$ before the boundary condition $\bphi$ is imposed. The path integral over $\bphi_B$ makes the resulting state fully mixed, except for the one condition $\bar\bphi_B=\vphi_B$.

A further simplification arises if we go to the momentum basis by defining
\be\label{tildepsi}
\tilde\Psi[t,J|K]= \int D\bphi D\bphi_{B}
e^{-i \int d^3\r (J(\r)\bphi(\r)+ K(\r)\bphi_B(\r))}
\Psi[t,\bphi|\bphi_B].
\ee
With this definition the canonical momenta are related to $J,K$ as $-i\frac{\delta}{\delta\bphi} = \sqrt{g_3} J$ and $-i\frac{\delta}{\delta\bphi_B}= \sqrt{g_3} K$, where
$\sqrt{g_3}$, the square root of the determinant of spatial metric, comes from the measure $d^3\r$ in \eqref{tildepsi}. In terms of $\tilde \Psi$,
\be\label{JLR}\begin{split}
p(t,\vphi|\vphi_B)=&\int \frac{dj  dk}{(2\pi)^2} e^{i(j\vphi+k\vphi_B)}
\int
DP DQ
\left(\tilde\Psi[t,J_L|K_L]\tilde\Psi^\dagger[t,J_R|K_R]
\right),\\[10pt]
J_{L/R}=& \frac{1}{2}j w  \pm P\\
K_{L/R}=& \frac{1}{2}k w  \pm Q.
\end{split}\ee
We would like to derive an equation for $\d_t p(t,\vphi|\vphi_B)$ instead of directly calculating $p(t,\vphi|\vphi_B)$. This is because the perturbation theory for calculating $p$ breaks down at large $t$ while the timescale for the evolution to become {\em Markovian} is expected to be $\O(1)$. The Markovian evolution has no explicit $t$ or $\vphi_B$ dependence and can be integrated for an arbitrarily long time. The time evolution of $p$ follows from the Schr\"odinger equation $i\d_t \Psi = H\Psi$, where
\be
H= \int dt d^3\r \sqrt{-g_{tt}} \left[\frac{1}{2} J^2+\frac{1}{2}|\nabla\phi|^2+V(\phi)\right].
\ee
\comment{\be
\d_t(\tilde\Psi[t,J_L(\r),K_L(\r)])\tilde\Psi^\dagger[t,J_R(\r),K_R(\r)])
= i (H\tilde\Psi[t,J_L(\r),K_L(\r)])\tilde\Psi^\dagger[t,J_R(\r),K_R(\r)]
-i\tilde\Psi[t,J_L(\r),K_L(\r)](H \tilde\Psi^\dagger[t,J_R(\r),K_R(\r)]).
\ee}
Only the momentum piece of the Hamiltonian survives the path integrals over $P,Q$ because they set $\bphi_L = \bphi_R$. 
So we get
\be\label{rhodot}
\d_tp(t,\vphi|\vphi_B) = -\d_\vphi \left(p(t,\vphi|\vphi_B) \expect{\bar P}_{\vphi,\vphi_B}\right)
\ee
where $\bar P= \int d^3\r \sqrt{-g_{tt}} w(r) P(\r)$ and 
\be
p(t,\vphi|\vphi_B)\expect{P(\r)}_{\vphi,\vphi_B}
=\int \frac{dj  dk}{(2\pi)^2} e^{i(j\vphi+k\vphi_B)}\int DP DQ \left(
\tilde\Psi[t,J_L|K_L]\tilde\Psi^\dagger[t,J_R|K_R]\right) P(\r),
\ee
where $J_{L/R},K_{L/R}$ are defined in \eqref{JLR}. 
\comment{
Using the Schr\"odinger equation
\be
i\d_t \Psi[t,\phi(\r)|\phi_B]= \int d^3\r [-\frac{1-r^2}{2} \frac{\delta^2}{\delta\phi(\r)^2}
+\frac{1}{2}(1-r^2)(\d_r\phi(\r))^2 +\frac{1}{2}(\nabla_\Omega\phi(\r))^2+ V(\phi(\r))]\Psi[t,\phi(\r)|\phi_B]
\ee
we find
\be\label{rhodot}
\d_tp(t,\vphi|\vphi_B)= \d_\vphi\int \frac{d J dJ_B}{(2\pi)^2} e^{iJ\vphi+iJ_B\vphi_B}
\int D\phi_B(\r)D\phi(\r)
e^{-i J\bar\phi -i J_B \bar\phi_B}
\frac{1}{V_0}\int_{r'<r_0} d^3\r' \Psi^\dagger[t,\phi(\r)|\phi_B(\r)]\overset{\leftrightarrow}{\d_{\phi(\r')}}
\Psi[t,\phi(\r)|\phi_B(\r)].
\ee}
\subsection{The free spectrum and low energy trunction}
To solve this path integral perturbatively in $V(\phi)$, we first need to diagonalize the free theory. For this purpose it is convenient to change the radial variable to $x$:
\be
x= \arctanh(r).
\ee
$x$ ranges from $0$ to $\infty$. In the new coordinates and after expanding $\phi$ in spherical harmonics the free field equation reads
\be
\d_t^2\phi_{l,m}(t,x) -\frac{1}{\tanh^2 x}\d_x(\tanh^2 x \d_x \phi_{l,m}(t,x))+\frac{l(l+1)}{\sinh^2 x} \phi_{l,m}(t,x)=0.
\ee
At large $x$ this becomes a free field on a (half) line. So we have a continuous spectrum:
\be
\phi(t,\r) = \sum_{l,m} \int_0^\infty \frac{d\omega}{2\pi} \phi_{\omega,l,m}(t)f_{\omega,l}(x)Y_{l,m}(\hat r),
\ee
where $f_{\omega,l}(x)$ is the eigensolution 
\be
\frac{1}{\tanh^2 x}\d_x(\tanh^2 x \d_x f_{\omega,l}(x))- \frac{l(l+1)}{\sinh^2 x} f_{\omega,l}(x)=-\omega^2 f_{\omega,l}(x),
\ee
that is regular at $x=0$, and normalized such that
\be
\int_0^\infty dx \tanh^2 x \ f_{\omega,l}(x) f_{\omega',l}(x) =2\pi \delta(\omega-\omega').
\ee
In particular, for $l=0$ and $\omega \ll 1$
\be\label{f0}
f_{\omega\ll 1,l=0}\simeq \frac{1}{\sqrt{\pi}} \cos(\omega x).
\ee
With this decomposition the free action diagonalizes into a sum of harmonic oscillators 
\be
S_0 = \frac{1}{2}\sum_{l,m}\int_0^\infty \frac{d\omega}{2\pi} \int dt [(\d_t\phi_{\omega,l,m})^2
-\omega^2 (\phi_{\omega,l,m})^2].
\ee
The potential $V(\phi)$ introduces coupling between various $\phi_{\omega,l,m}$. However, this interaction is localized near the origin $x\sim 1$:
\be\label{sint}
S_{\rm int} = -\int dt dx d\hat r \frac{\tanh^2x}{\cosh^2 x} V(\phi).
\ee
We are interested in the late time behavior of the system. This is controlled by the low frequency modes $\omega\ll 1$. Because of the centrifugal barrier low frequency modes with $l\neq 0$ are suppressed in the interaction region:
\be
f_{\omega,l}(x\ll \omega^{-1}) \sim (\omega x)^l.
\ee
Therefore the leading effect of interactions can be understood by truncating to $l=0$ sector. These modes have a nearly constant profile in the interaction region $f_{\omega\ll 1,l=0}(x\ll \omega^{-1}) \simeq 1/\sqrt{\pi}$. So in the interaction term, we can replace
\be\label{phiLamb}
\phi \to \int_0^\Lambda \frac{d\omega}{2\pi^{3/2}} \phi_{\omega,0,0},
\ee
where we introduced a cutoff $\Lambda\ll 1$ up to which the approximation \eqref{phiLamb} is reliable. Substituting this in \eqref{sint} gives an all-to-all interaction among the $l=0$ low frequency modes:
\be\label{all}
S_{\rm int}\simeq -\frac{4\pi}{3} \int dt\ V\left(\int_0^\Lambda \frac{d\omega}{2\pi^{3/2}} \phi_{\omega,0,0}(t)\right).
\ee
On the other hand, the observable $\bar\phi$ is also a superposition of $l=0$ modes
\be
\bar\phi(t) = 4\pi \int dx \frac{\tanh^2 x}{\cosh x} w(x) 
\int_0^\infty \frac{d\omega}{2\pi}\phi_{\omega,0,0}(t) f_{\omega,0}(x).
\ee
As discussed above, we would like the characteristic size of $w$ to be comparable to the horizon size. In $x$ coordinates this corresponds to a size $x_0=\arctanh(r_0)> 1$. Hence the profile \eqref{f0} implies that all modes with $\omega x_0 \ll 1$ contribute equally to $\bar\phi$. That is, for low-frequency modes we can approximate
\be\label{phibarL}
\bar\phi(t) \simeq \int_0^{1/x_0} \frac{d\omega}{2\pi^{3/2}}\phi_{\omega,0,0}(t),
\ee
which coincides with \eqref{phiLamb} except for the replacement $\Lambda\to 1/x_0$. The approximate classicality of $\bar\phi$ can also be understood from the fact that it is composed of $\omega\ll 1$ modes which have large occupation number at dS temperature. 

When $t\gg x_0$, the $\omega$ integrals in the leading low-frequency approximation are dominated at 
\be
\omega \sim \frac{1}{t} \ll \frac{1}{x_0}\ll \Lambda.
\ee
Therefore, the difference of the upper bounds in \eqref{phiLamb} and \eqref{phibarL} becomes unimportant and they can be sent to $\infty$.


The same substitutions can be made in the source terms. Using 
\be
J_{L}(\r) =\frac{1}{2} j w(r)+P(\r)
\ee
and decomposing
\be
P(\r) = \sum_{l,m}\int \frac{d\omega}{2\pi \sqrt{-g_{tt}}} P_{\omega,l,m} f_{\omega,l}(x)Y_{l,m}(\hat r),
\ee
we get
\be\label{j}
\int d^3\r J_L(\r) \bphi_L(\r) \simeq  \int \frac{d\omega}{2\pi} \left(\frac{j}{\sqrt{\pi}}+P_{\omega,0,0}\right) 
\bphi_{L,\omega,0,0},
\ee
with a similar expression for $\bphi_R$ and for $\bphi_{B,L/R}$. Finally $\bar P$ in \eqref{rhodot} simplifies to
\be\label{Pbar}
\int d^3\r \sqrt{-g_{tt}} w(r) P(\r) \simeq \int \frac{d\omega}{2\pi^{3/2}} P_{\omega,0,0}.
\ee
\subsection{Free field diffusion}\label{sec:diff}
In the free theory, all integrals in \eqref{p} become Gaussian and can be evaluated to give
\be\label{random}
p_0(t,\vphi|\vphi_B)
=\sqrt{\frac{2\pi}{t}}e^{- \frac{2\pi^2}{t}(\vphi-\vphi_B)^2}.
\ee
This describes how the free massless field diffuses in de Sitter. The attractive potential will eventually contain this spread. However for a generic $V$ the path integral can be calculated only perturbatively. To set up the perturbation theory, we need various free partition functions (or wavefunctions) as functions of the sources. These will be calculated in detail in the appendix. 

As discussed earlier, it makes sense to calculate $\d_t p$ rather than $p$ itself. Let us see how this works in the free case. We will find in appendix \ref{app:Z} that the free partition function is localized at $j=-k$:
\be\label{deltaj+}
Z_0(t,j|k)\propto \delta(j_+),
\ee
where
\be\label{Jpm}
j_{\pm} \equiv \frac{j \pm k}{2}.
\ee
This is to be expected since for a shift-symmetric field $p_0(t,\vphi|\vphi_B)$ is only a function $\vphi-\vphi_B$. The average momentum $\bar P$ as a function of the sources is found in \eqref{Pk} to be
\be
\bar P =-\frac{i}{8\pi^2} k = \frac{i}{8\pi^2} (j-2j_+).
\ee
Therefore, we have from \eqref{rhodot}
\be\begin{split}
\d_t p_0(t,\vphi|\vphi_B)=&
\d_\vphi\int \frac{dj dk}{(2\pi)^2} e^{i(j\vphi+k\vphi_B)} \frac{i}{8\pi^2} (j-2j_+) Z(t,j|k)\\[10pt]
=& \frac{1}{8\pi^2}\d_\vphi^2 p_0(t,\vphi|\vphi_B),
\end{split}\ee
where in the second line we discarded the $j_+$ term because of \eqref{deltaj+}. This is the diffusion equation whose normalized solution is \eqref{random}.
\subsection{Interacting field}
As mentioned before, perturbation theory for calculating $p(t,\vphi|\vphi_B)$ breaks down at long times. This can be clearly seen from the fact that $p$ has to evolve from the random-walk distribution \eqref{random} at short times to the equilibrium distribution \eqref{eq2}. Hence, ultimately, the interaction Hamiltonian becomes more important than the free part. However for a shallow potential (as characterized in \eqref{sr}) the transition time $t_V$ is long. So we can treat the potential perturbatively for  $1\ll t\ll t_V$. Our strategy is to show that in this regime the evolution of $p(t,\vphi|\vphi_B)$ becomes Markovian, namely explicit dependence on $t$ and $\vphi_B$ disappears. Once this is the case one can integrate this Markovian evolution for an arbitrarily long time. 

The ingredients for the perturbation theory are the vertices which couple the free modes, and the propagators. In the low energy limit, relevant for $t\gg 1$, these elements simplify. The potential introduces a vertex that couples all $l=0$ infrared modes together \eqref{all}. The propagators are worked out in the appendix. Since we are setting boundary conditions, there are external lines. They are found in appendix \ref{app:ext} to be 
\be
\expect{\bar\phi_r(t')}^{(0)} = \vphi_B + \frac{t'}{t}(\vphi-\vphi_B),
\ee
\be
\expect{\bar\phi_i(t')}^{(0)} = i\frac{\pi}{2t}(\vphi-\vphi_B),
\ee
where $(0)$ stands for zeroth order in $V$, and $\phi_{r/i}$ are defined in terms of the fields coming from the two sides of the path integral contour
\be
\phi_{r/i} = \frac{\phi_L\pm \phi_R}{2}.
\ee
$\phi_{\omega,i}$ is suppressed by an extra power of $\omega$ with respect to $\phi_{\omega,r}$. This explains, via \eqref{phibarL}, the suppression of $\bar\phi_i(t')$ at long times. Therefore the leading low-frequency approximation corresponds to keeping the minimum number of $\phi_i$ fields. Given that every new interaction vertex comes with a time integral and a factor of $i$, there has to be at least one $\phi_i$ per vertex. 

The internal propagators are found in appendix \ref{app:int} to be 
\be\label{pp}
\expect{\bar\phi_r(t') \bar\phi_r(t'')}^{(0)_c}_{t'>t''}= \frac{t'' (t-t')}{4\pi^2 t},
\ee
\be\label{pm}
\expect{\phi_r(t')\phi_i(t'') }^{(0)_c}= \frac{i }{8\pi t}(t' -t \theta(t'-t'') + t\theta(t''-t')),
\ee
where $(0)_c$ stands for zeroth order in $V$ and connected. $\expect{\phi_r\phi_r}^{(0)_c}$ is not relevant at leading order as will be seen shortly. Note that at $t'=t$ the internal propagators vanish and $\expect{\bar\phi_r(t)}^{(0)}= \vphi$. 

We will also need the correlator of the momentum operator $P$ with the fields in order to evaluate $\expect{\bar P}_{\vphi,\vphi_B}$ in \eqref{rhodot}. Since we are perturbing around the free theory, we subtract the piece that gives the free diffusion equation:
\be
\bar P'= \bar P-\frac{i}{8\pi^2}j.
\ee
It is shown in appendix \ref{app:int} that
\be\label{Pi}
\expect{\bar P' \bar\phi_i(t')}^{(0)_c} =\frac{i}{4\pi}\delta(t-t'),
\ee
and $\expect{\bar P'\phi_r(t')}^{(0)_c}=0$. 

The non-analytic time-dependence of the above expressions is unphysical. It is an artifact of sending the frequency cutoff to infinity by sending $x_0\to 0$ in the low frequency approximation. For a finite $x_0$, sharp changes become smooth. In particular $\expect{\bar\phi_i(t')}^{(0)}$ goes to zero as $t'\to t$, and \eqref{Pi} becomes a smeared delta function.

Using the above propagators, the first order correction to $\d_t p$ arises from bringing down one interaction from $e^{i(S_{{\rm int},L}-S_{{\rm int},R})}$ and using the low energy approximation \eqref{all} for $S_{\rm int}$:
\be\label{1st}
\expect{\bar P' }^{(1)_c}_{\vphi,\vphi_B}
=-\frac{8\pi i}{3}\int_0^t dt' \expect{\bar P \bar\phi_i(t')}^{(0)_c} V'\left(\expect{\bar\phi_r(t')}^{(0)}\right)
=\frac{1}{3}V'(\vphi).
\ee
Note that because of the upper bound $t$ on the $t'$ integral only half of the delta function in \eqref{Pi} contributes. 

Now consider a connected diagram at higher order in perturbation theory. It must contain a vertex like \eqref{1st}, but with at least one other field being internal. Since there is already one $\bar\phi_i$ contracted with $\bar P'$, this other field has to be $\phi_r$ at leading low-frequency limit. Since $\expect{\bar P \bar\phi_i(t')}^{(0)_c}\propto \delta(t-t')$ and the propagators \eqref{pp}, \eqref{pm} vanish at $t'= t$ all such connected diagrams vanish. Hence at leading low energy limit, but to all orders in $V$,
\be\label{FP1}
\d_t p(t,\vphi|\vphi_B)= \frac{1}{8\pi^2} \d_\vphi^2 p(t,\vphi|\vphi_B)
+\frac{1}{3} \d_\vphi(V'(\vphi)p(t,\vphi|\vphi_B)).
\ee
\section{Discussion}\label{sec:con}
We studied the equilibrium state and the out of equilibrium dynamics of a scalar field in the static patch of de Sitter. Because of the presence of the cosmological horizon the equilibrium state is a mixed thermal state. When further reduced to a small subspace, like the spatial average of the field over the region $r<r_0$, any perturbed state is expected to relax and equilibrate. In order to study this relaxation, we set up a perturbative scheme to calculate the evolution of $p(t,\vphi|\vphi_B)$, the diagonal element of the reduced density matrix with a particular boundary condition imposed at $t=0$.

Our perturbative method, which is based on diagonalizing the free static patch Hamiltonian, does not use the full symmetry of de Sitter spacetime. However, it is well-suited to study the long time behavior of the system via a low energy trunction. In this limit, we found the well-known Fokker-Planck equation \eqref{FP1} for $p(t,\vphi|\vphi_B)$. From this equation the equilibrium distribution \eqref{eq}, and the exponents that control the late-time approach to equilibrium can be derived following \cite{SY}. For instance, the slowest decaying mode was found in \cite{SY} to have a frequency $ i\omega \approx 1.4 \sqrt{\lambda/24\pi^2}$ for $V=\lambda\phi^4/4$. The non-perturbative nature of the result is manifest in the non-analytic dependence on $\lambda$.

However, in our setup it is possible (at least in principle) to systematically include higher order corrections to \eqref{FP1} by keeping the subleading terms in the low-frequency expansion. An important question that was left unanswered is whether the evolution continues to be Markovian beyond the leading order. This is not manifest in our approach, but it is expected to be the case. The finite temperature $T=1/2\pi$ sets a natural scale for the non-locality in time. At longer times one would expect a derivative expansion to be applicable. 

Finally, while it is automatic in the leading low energy approximation that long-time observables (such as the relaxation exponents) are independent of $r_0$, at subleading orders it is not obvious from our formalism. We have checked this at next-to-leading order and for a few implementations of the cutoff. We found that with an appropriate field redefinition the evolution equation, and hence the exponents, remain unchanged.\footnote{We thank Victor Gorbenko for suggesting this to us.} We leave further exploration of these points to future work \cite{memory}.

\comment{
In particular, the invariance of the evolution equation implies that the equilibrium distribution has to be invariant when expressed in terms of the renormalized field $\tilde \vphi$. Making explicit the $r_0$ dependence via the weight function and using the transformation law of a distribution $p(r_0,\vphi) d\vphi = \tilde p(r_0,\tilde \vphi) d\tilde\vphi$ then implies
\be
\d_{r_0} \tilde p_{HH}(r_0,\tilde\vphi) + \d_{\tilde\vphi}(\tilde p_{HH}(r_0,\tilde\vphi) \d_{r_0} \tilde\vphi)=0.
\ee
From \eqref{reduce} with $\vphi_L=\vphi_R=\vphi$, we find that 
\be\label{ren}
\d_{r_0} \tilde\vphi=\expect{\int {d^3\r}\d_{r_0}w(r)\phi(\r)}_{HH,\vphi}
\ee
where the $r_0$ dependence of the weight function $w$ is implicit, and $\expect{}_{HH,\vphi}$ is defined in terms of the projection $\mathcal{P}_\vphi$ into the $\bar\phi = \vphi$ subspace:
\be
\expect{O}_{HH,\vphi} = \tr[O \mathcal{P}_{\vphi} \rho_{HH}].
\ee
From \eqref{ren} follows that $\vphi-\tilde\vphi = \O(V')$. Hence the corrections introduced by this field-redefinition into the equilibrium state \eqref{eq2} are ``slow-roll'' suppressed.
}

\section*{Acknowledgments}
We thank Victor Gorbenko, Diana L\'opez Nacir, Kyriakos Papadodimas and Sergey Sibiryakov for stimulating discussions. This work was partially supported by the Simons Foundation Origins of the Universe program (Modern Inflationary Cosmology collaboration).
\appendix
\section{Perturbation Theory}
To calculate $p(t,\vphi|\vphi_B)$ or its time evolution perturbatively in $V(\phi)$ we follow the standard procedure. At every order one draws an interaction vertex from $e^{\pm iS_{\rm int}}$ coming from $\Psi$ or $\Psi^\dagger$ and contracts fields using the free partition function. The difference with other familiar examples is that the path integral is performed in several steps. First there are two path integrals (L and R) from $0 \to \mp i\pi+t$ for $\tilde \Psi$ and $\tilde\Psi^\dagger$. We introduced the sources $J_{L/R},K_{L/R}$ for the boundary conditions of these path integrals (i.e. they are the conjugate momenta to the boundary values $\bphi,\bphi_B$). Then the path integral over the momenta $P =(J_L - J_R)/2$ and $Q=(K_L-K_R)/2$ identifies the boundary conditions of the L and R paths. This is relevant for the discussion of the diagonal element of the density matrix. The sources for the boundary configurations $\bphi,\bphi_B$ are $J_L+J_R$ and $K_L+K_R$ respectively. They are identified as
\be\label{j1}
J_L(\r)+J_R(\r) = j w(r),\qquad K_L(\r)+K_R(\r) = k w(r)
\ee
and $j,k$ are integrated over as in \eqref{p} to reduce the density matrix to a single observable with boundary values $\vphi, \vphi_B$. Below we will first calculate the free partition functions with arbitrary sources. Using that we find the external lines and internal propagators as the necessary ingredients of the perturbative expansion.
\subsection{Free partition function}\label{app:Z}
We diagonalize the free theory in terms of $\phi_{\omega,l,m}$ modes. The free wavefunction $\tilde \Psi_0$ with arbitrary sources $S_{\omega,l,m}$ factorizes. So for every mode we have
\be
\tilde\Psi_{0,\omega,l,m}=\int D\phi_{\omega,l,m}e^{-\int_0^{u_1} du \left[\frac{1}{2}(\dot\phi_{\omega,l,m}^2+\omega^2 \phi_{\omega,l,m}^2)
+i\phi_{\omega,l,m}S_{\omega,l,m}\right]}
\ee
where we used Euclidean time. Eventually $u_1$ is analytically continued to $\pi+it$. The solution can be written in terms of the Neumann Green's function
\be
G(u,v)=-\frac{\cosh(\omega (u_1 -u))\cosh(\omega v)}{\omega \sinh(\omega u_1)},\qquad u>v
\ee
and the same expression with $u\leftrightarrow v$ when $u<v$. In terms of this
\be\label{Zpsi}
\tilde\Psi_{0,\omega,l,m} = \exp\left[\frac{1}{2}\int_{0}^{u_1}du \int_{0}^{u_1} d v G(u,v) S_{\omega,l,m}(u) S_{\omega,l,m}(v)\right] .
\ee
This expression will give the internal (L) propagator between two arbitrary times $u$ and $v$ when interactions are included. Once the propagators are derived we set $S=0$ except at $u=0$ and $u=u_1= \pi+i t$, where it is identified with $K_L$ and $J_L$, respectively. The resulting wavefunction function is (we only consider $l=m=0$, which are most relevant at late times, and drop the corresponding indices)
\be\label{psi}
\log(\tilde\Psi_{0,\omega})=-\frac{2 J_{L,\omega} K_{L,\omega}+(J_{L,\omega}^2+K_{L,\omega}^2)\cosh(\omega (\pi+it))}
{2\omega \sinh(\omega (\pi+it))}, 
\ee
$\tilde\Psi^\dagger_\omega(t,J_R|K_R)$ is given by a similar expression with $L\leftrightarrow R, t\to -t$. 

Next we multiply $\tilde\Psi_\omega(t,J_L|K_L)\tilde\Psi_\omega^\dagger(t,J_R|K_R)$ and substitute
\be\label{JL/R}
J_{L/R,\omega}= \frac{1}{2} J_\omega \pm P_\omega,\qquad K_{L/R,\omega}=\frac{1}{2} K_{\omega} \pm Q_\omega.
\ee
Performing the path integral over $P,Q$ identifies the boundary conditions on the left and right branch. The path integral over $P_\omega,Q_\omega$ is a Gaussian in the free theory, with the saddle at 
\be\label{P,Q}
P_\omega = -i\frac{\sin(\omega t)}{2\sinh(\omega \pi)}K_\omega,\quad
Q_\omega =  -i\frac{\sin(\omega t)}{2\sinh(\omega \pi)}J_\omega.
\ee
After the identification \eqref{j1} and taking the low-frequency limit we find
\be\label{J,K}
J_\omega = \frac{j}{\sqrt{\pi}},\qquad K_\omega=\frac{k}{\sqrt{\pi}}.
\ee
The average $\bar P$ in the low-frequency approximation \eqref{Pbar} would then become
\be\label{Pk}
\bar P \simeq -i k\int \frac{d\omega}{4\pi^3}\frac{\sin(\omega t)}{\omega}= -\frac{i}{8\pi^2}k.
\ee
To find the partition function $Z(t,j|k)$, we substitute the saddle solution \eqref{P,Q} in $\tilde\Psi_\omega\tilde\Psi_\omega^\dagger$ and take the limit $\omega\to 0$ at fixed $\omega t $. We obtain
\be
\log Z_\omega =-\frac{\cos^2(\omega t/2)}{\pi^2 \omega^2} j_{+}^2
-\frac{\sin^2(\omega t/2)}{\pi^2 \omega^2} j_{-}^2,
\ee
where we defined $j_{\pm} \equiv \frac{j \pm k}{2}$. Dropping the normalization, the free partition function \eqref{Z} is 
\be\label{Zjk}
Z_0(t,j|k)\propto\exp\left(\int \frac{d\omega}{2\pi}\log Z_\omega\right).
\ee
The integration over the coefficient of $j_+^2$ is infrared singular. Therefore $Z_0(t,j|k)\propto \delta(j_+)$ which implies that $p(t,\vphi|\vphi_B)$ depends on the fields only via $\vphi-\vphi_B$. Performing the other integral over the coefficient of $j_-^2$ gives (up to a normalization factor $N$)
\be
Z_0(t,j|k)=N e^{-\frac{t}{8\pi^2}j_-^2}\delta(j_+).
\ee
Substitution in \eqref{p} and normalizing the distribution gives the solution to the diffusion equation \eqref{random}.
\subsection{External lines}\label{app:ext}
Since there are nonzero external sources in the $\tilde \Psi$ path integral, there can be external lines in this part of the path integral:
\be
\phi_{L,\omega}(t') = 
-\frac{i}{\omega\sinh(\omega (\pi+i t))}(J_{\omega,L} \cosh(\omega (\pi+it'))+K_{\omega,L} \cos(\omega(t-t'))),
\ee
with $L\to R$ and $t\to -t,t'\to -t'$ for $\tilde\Psi^\dagger$. It is useful to combine the left and right fields and define
\be
\phi_{r/i,\omega}(t')= \frac{1}{2}(\phi_{\omega,L}(t')\pm\phi_{\omega,R}(t')).
\ee
Next we need to substitute \eqref{JL/R} and perform the integral over $Q_\omega,P_\omega$. To find the external lines of this path integral, we insert the solution \eqref{P,Q} for $P,Q$ and make the identification \eqref{J,K}. Taking the low energy limit we find
\be\label{phir}
\phi_{r,\omega}(t') = -i \frac{j\cos(\omega(t-t'))+k\cos(\omega t')}{2\pi^{3/2}\omega^2}
\ee
\be\label{phii}
\phi_{i,\omega}(t')= -\frac{j\sin(\omega(t-t'))}{2\pi^{1/2}\omega}.
\ee
We integrate these external lines over $\omega$ to obtain $\bar\phi(t')$, the combination that (at low frequencies) enters the all-to-all interaction \eqref{all}. We see that the coefficient of $j_+=(j+k)/2$ in $\phi_r$ integral diverges. We can relate this to the divergent coefficient of $j_+^2$ in the free partition function \eqref{Zjk}. The coefficient of $j_-$ in $\bar\phi_r(t')$ is $i(t-2t')/8\pi^2$. So we can write
\be\label{phir-}
\int \frac{dj dk}{(2\pi)^2} e^{i(j\vphi+k\vphi_B)}\bar\phi_r(t') Z_0(t,j|k)
=\int \frac{dj dk}{(2\pi)^2} e^{i(j\vphi+k\vphi_B)}
\left(\frac{i}{2}\d_{j_+}+\frac{t-2t'}{8\pi^2} \overset{\leftarrow}{\d_{\vphi_-}}\right)
  Z_0(t,j|k),
\ee
where $\vphi_\pm = \vphi\pm \vphi_B$. $\d_{\vphi_-}$ can be moved outside to act on $p_0$, the free distribution \eqref{random}, and $\d_{j_+}$ be integrated by parts to give the external field
\be\label{phirfree}
\expect{\bar\phi_r(t')}^{(0)} = \vphi_B + \frac{t'}{t}(\vphi-\vphi_B).
\ee
Similarly in the case of $\phi_i$ we get
\be\label{phii2}
\expect{\bar\phi_i(t')}^{(0)} =\frac{i}{8\pi} \frac{\d}{\d\vphi}\log p_0,
\ee
which gives
\be\label{phiifree}
\expect{\bar\phi_i(t')}^{(0)}= -i\frac{\pi}{2t}\vphi_- 
= -i \frac{\pi}{2} \frac{d}{dt'}\expect{\bar\phi_r(t')}^{(0)}.
\ee
\subsection{Internal lines}\label{app:int}
Internal lines can come from (a) contracting two $\phi_{L,\omega}$ in $\tilde\Psi$ path integral (or two $\phi_{R,\omega}$ in $\tilde\Psi^\dagger$), (b) contracting $\{P_\omega,Q_\omega\}$ in the external fields to the first path integral in the momentum path integral, or (c) contracting two fields in the final integration over ${j,k}$.

The internal propagator is derived from \eqref{Zpsi}
\be\label{Gint}
\expect{\phi_{L,\omega_1}(t')\phi_{L,\omega_2}(t'')}= \frac{\cosh(\omega_1(\pi+i t''))\cos(\omega_1(t-t'))}
{\omega_1 \sinh(\omega_1(\pi+it))} 2\pi\delta(\omega_1-\omega_2).
\ee
Next, every external $\phi_{\omega,L}$ in $\tilde\Psi$ path integral contains a momentum contribution
\be
\left.\phi_{L,\omega}(t')\right|_{j=k=0} = -i\frac{Q(\omega)\cos(\omega(t-t'))+P(\omega)\cosh(\omega(\pi+it'))}
{\omega\sinh(\omega(\pi+it))}
\ee
with $\phi_{\omega,R}(t')$ given by the complex conjugate of this. These can be contracted in $P,Q$ path integral. The explicit form of the Gaussian integrand can be obtained by substituting \eqref{JL/R} in  \eqref{psi} and its complex conjugate:
\be
\left.\log(\tilde\Psi[t,J_L|K_L]\tilde\Psi^\dagger[t,J_R|K_R])\right|_{j=k=0}=-\sinh(\omega \pi)\frac{(P^2(\omega)+Q^2(\omega))\cosh(\omega\pi)+2 P(\omega)Q(\omega)\cos(\omega t)}{\pi \omega |\sinh(\omega(\pi-i t))|^2}.
\ee
Adding the resulting correlators to \eqref{Gint} gives at leading order in small $\omega$
\be\label{rr}
\expect{\phi_{\omega,r}(t')\phi_{\omega,r}(t'')}^{(0)}_{(a),(b)}=\frac{\cos(\omega(t''-t'))}{2\pi\omega^2},
\ee
\be\label{ri}
\expect{\phi_{\omega,r}(t')\phi_{\omega,i}(t'')}^{(0)}_{(a),(b)}=\frac{i \sin(\omega(t''-t'))}{2\omega}.
\ee
Note that all poles in \eqref{Gint} canceled except the one at $\omega =0$. 

As before we integrate over $\omega$. The double pole in \eqref{rr} gives a singularity. However there is another contribution (c) to the internal lines from contraction of $\bar\phi$ fields in the ${j,k}$ integral:
\be\begin{split}
&\int \frac{dj dk}{(2\pi)^2} e^{i(j\vphi+k\vphi_B)}\bar\phi_r(t')\bar\phi_r(t'') Z_0(t,j|k)\\[10pt]
&=\int \frac{dj dk}{(2\pi)^2} e^{i(j\vphi+k\vphi_B)}
\left[\left(\frac{i}{2}\d_{j_+}+\frac{t-2t'}{8\pi^2} \overset{\leftarrow}{\d_{\vphi_-}}\right)
\left(\frac{i}{2}\d_{j_+}+\frac{t-2t''}{8\pi^2} \overset{\leftarrow}{\d_{\vphi_-}}\right)
-\int \frac{d\omega}{2\pi}\frac{1+\cos(\omega t)}{4\pi^2 \omega^2}\right]Z_0(t,j|k)\\[10pt]
&=\left[\expect{\phi_r(t')}^{(0)}\expect{\phi_r(t')}^{(0)}
-\int \frac{d\omega}{2\pi}\frac{1+\cos(\omega t)}{4\pi^2 \omega^2}
-\frac{(t-2t')(t-2 t'')}{16\pi^2 t}\right]p_0(t,\vphi|\vphi_B).\end{split}\ee
The last two terms once summed with \eqref{rr} cancel the infrared singularity and give the full propagator:
\be\label{plus}
\expect{\bar\phi_r(t') \bar\phi_r(t'')}^{(0)_c}_{t''<t'}= \frac{t'' (t-t')}{4\pi^2 t},
\ee
where the superscript $(0)_c$ denotes zeroth order in $V$ and connected.
\comment{
Note that this could be obtained by expressing $\phi_r$ in terms of $\vphi$ and its conjugate momentum:
\be
\phi_r(t')\to \vphi +\frac{(t-t')}{4\pi^2} \d_\vphi
\ee
and $\phi_r(t'')$ by \eqref{phirfree}, and time-ordering. }
Similarly, we find
\be
\expect{\bar\phi_r(t')\bar\phi_i(t'') }^{(0)_c}= \frac{i }{8\pi t}(t' -t \theta(t'-t'') + t\theta(t''-t')).
\ee
Note that these propagators vanish when $t'=t$. 

Finally in the evaluation of $\d_t p(t,\vphi|\vphi_B)$ as in \eqref{rhodot} we need to correlate $P$ with the fields in the interaction vertex $V(\phi)$. The momentum contraction at low energy simplifies to 
\be\label{Pii}
\expect{P_\omega \phi_{\omega,i}(t')}^{(0)_c}_{j=0,k =0} = \frac{i}{2} \cos(\omega(t-t')).
\ee
\be\label{Pr}
\expect{P_\omega \phi_{\omega,r}(t')}^{(0)_c}_{j=0,k =0} = \frac{\sin(\omega(t-t'))}{2\omega\pi}.
\ee
However when there are sources there is an external $\bar P$ in the $P,Q$ path integral given by $-ik/8\pi^2$ \eqref{Pk}. We subtract from this the part that gives the free diffusion equation as in section \ref{sec:diff}:
\be
\bar P' = \bar P-\frac{i}{8\pi^2} j = -\frac{i}{4\pi^2} j_+ +\text{internal to $P,Q$ integral}.
\ee
The $j_+$ piece can be contracted with $\phi_r$ fields coming from the interaction vertices:
\be\begin{split}
&\int \frac{dj dk}{(2\pi)^2} e^{i(j\vphi+k\vphi_B)} j_+ \bar\phi_r(t') Z_0(t,j|k)\\[10pt]
&=\int \frac{dj dk}{(2\pi)^2} e^{i(j\vphi+k\vphi_B)} j_+
\left(\frac{i}{2}\d_{j_+}+\frac{t-2t'}{8\pi^2} \overset{\leftarrow}{\d_{\vphi_-}}\right)
Z_0(t,j|k)\\[10pt]
&=-\frac{i}{2} p_0(t,\vphi|\vphi_B).\end{split}\ee
Integrating \eqref{Pii} and \eqref{Pr} over $\omega$ and adding the above $j_+$ contraction gives 
\be
\expect{\bar P' \bar\phi_r}^{(0)_c}=0,
\ee
\be
\expect{\bar P' \bar\phi_i(t')}^{(0)_c} =\frac{i}{4\pi}\delta(t-t').
\ee
\bibliography{bibstat}

\providecommand{\href}[2]{#2}\begingroup\raggedright\begin{thebibliography}{10}

\bibitem{Starobinsky}
A.~A. Starobinsky, ``{STOCHASTIC DE SITTER (INFLATIONARY) STAGE IN THE EARLY
  UNIVERSE},''
\href{http://dx.doi.org/10.1007/3-540-16452-9_6}{{\em Lect. Notes Phys.}
  {\bfseries 246} (1986) 107--126}.

\bibitem{SY}
A.~A. Starobinsky and J.~Yokoyama, ``{Equilibrium state of a selfinteracting
  scalar field in the De Sitter background},''
  \href{http://dx.doi.org/10.1103/PhysRevD.50.6357}{{\em Phys. Rev.} {\bfseries
  D50} (1994) 6357--6368},
\href{http://arxiv.org/abs/astro-ph/9407016}{{\ttfamily arXiv:astro-ph/9407016
  [astro-ph]}}.

\bibitem{Tsamis}
N.~C. Tsamis and R.~P. Woodard, ``{Stochastic quantum gravitational
  inflation},'' \href{http://dx.doi.org/10.1016/j.nuclphysb.2005.06.031}{{\em
  Nucl. Phys.} {\bfseries B724} (2005) 295--328},
\href{http://arxiv.org/abs/gr-qc/0505115}{{\ttfamily arXiv:gr-qc/0505115
  [gr-qc]}}.

\bibitem{Finelli}
F.~Finelli, G.~Marozzi, A.~Starobinsky, G.~Vacca, and G.~Venturi, ``{Generation
  of fluctuations during inflation: Comparison of stochastic and
  field-theoretic approaches},''
  \href{http://dx.doi.org/10.1103/PhysRevD.79.044007}{{\em Phys. Rev. D}
  {\bfseries 79} (2009) 044007},
  \href{http://arxiv.org/abs/0808.1786}{{\ttfamily arXiv:0808.1786 [hep-th]}}.

\bibitem{Finelli2}
F.~Finelli, G.~Marozzi, A.~Starobinsky, G.~Vacca, and G.~Venturi, ``{Stochastic
  growth of quantum fluctuations during slow-roll inflation},''
  \href{http://dx.doi.org/10.1103/PhysRevD.82.064020}{{\em Phys. Rev. D}
  {\bfseries 82} (2010) 064020},
  \href{http://arxiv.org/abs/1003.1327}{{\ttfamily arXiv:1003.1327 [hep-th]}}.

\bibitem{Burgess}
C.~P. Burgess, R.~Holman, and G.~Tasinato, ``{Open EFTs, IR effects \&
  late-time resummations: systematic corrections in stochastic inflation},''
  \href{http://dx.doi.org/10.1007/JHEP01(2016)153}{{\em JHEP} {\bfseries 01}
  (2016) 153},
\href{http://arxiv.org/abs/1512.00169}{{\ttfamily arXiv:1512.00169 [gr-qc]}}.

\bibitem{Vennin}
V.~Vennin and A.~A. Starobinsky, ``{Correlation Functions in Stochastic
  Inflation},'' \href{http://dx.doi.org/10.1140/epjc/s10052-015-3643-y}{{\em
  Eur. Phys. J.} {\bfseries C75} (2015) 413},
\href{http://arxiv.org/abs/1506.04732}{{\ttfamily arXiv:1506.04732 [hep-th]}}.

\bibitem{Kitamoto}
H.~Kitamoto, ``{Infrared resummation for derivative interactions in de Sitter
  space},'' \href{http://dx.doi.org/10.1103/PhysRevD.100.025020}{{\em Phys.
  Rev. D} {\bfseries 100} no.~2, (2019) 025020},
  \href{http://arxiv.org/abs/1811.01830}{{\ttfamily arXiv:1811.01830
  [hep-th]}}.

\bibitem{Markkanen}
T.~Markkanen, A.~Rajantie, S.~Stopyra, and T.~Tenkanen, ``{Scalar correlation
  functions in de Sitter space from the stochastic spectral expansion},''
  \href{http://dx.doi.org/10.1088/1475-7516/2019/08/001}{{\em JCAP} {\bfseries
  08} (2019) 001}, \href{http://arxiv.org/abs/1904.11917}{{\ttfamily
  arXiv:1904.11917 [gr-qc]}}.

\bibitem{Lopez}
D.~López~Nacir, F.~D. Mazzitelli, and L.~G. Trombetta, ``{To the sphere and
  back again: de Sitter infrared correlators at NTLO in 1/N},''
  \href{http://dx.doi.org/10.1007/JHEP08(2019)052}{{\em JHEP} {\bfseries 08}
  (2019) 052},
\href{http://arxiv.org/abs/1905.03665}{{\ttfamily arXiv:1905.03665 [hep-th]}}.

\bibitem{Gorbenko}
V.~Gorbenko and L.~Senatore, ``{$\lambda \phi^4$ in dS},''
\href{http://arxiv.org/abs/1911.00022}{{\ttfamily arXiv:1911.00022 [hep-th]}}.

\bibitem{Marolf}
D.~Marolf and I.~A. Morrison, ``{The IR stability of de Sitter: Loop
  corrections to scalar propagators},''
  \href{http://dx.doi.org/10.1103/PhysRevD.82.105032}{{\em Phys. Rev.}
  {\bfseries D82} (2010) 105032},
\href{http://arxiv.org/abs/1006.0035}{{\ttfamily arXiv:1006.0035 [gr-qc]}}.

\bibitem{HH}
J.~B. Hartle and S.~W. Hawking, ``{Wave Function of the Universe},''
  \href{http://dx.doi.org/10.1103/PhysRevD.28.2960}{{\em Phys. Rev.} {\bfseries
  D28} (1983) 2960--2975}.
[Adv. Ser. Astrophys. Cosmol.3,174(1987)].

\bibitem{Higuchi}
A.~Higuchi, D.~Marolf, and I.~A. Morrison, ``{On the Equivalence between
  Euclidean and In-In Formalisms in de Sitter QFT},''
  \href{http://dx.doi.org/10.1103/PhysRevD.83.084029}{{\em Phys. Rev.}
  {\bfseries D83} (2011) 084029},
\href{http://arxiv.org/abs/1012.3415}{{\ttfamily arXiv:1012.3415 [gr-qc]}}.

\bibitem{GH}
G.~W. Gibbons and S.~W. Hawking, ``{Cosmological Event Horizons,
  Thermodynamics, and Particle Creation},''
\href{http://dx.doi.org/10.1103/PhysRevD.15.2738}{{\em Phys. Rev.} {\bfseries
  D15} (1977) 2738--2751}.

\bibitem{memory}
M.~Mirbabayi, ``Loss of Memory in de Sitter''~{\it in preparation.}

\end{thebibliography}\endgroup

\end{document}